\documentclass[journal,onecolumn,11pt]{IEEEtran}
\usepackage{amsmath,amssymb,mathtools,amsthm}
\usepackage{booktabs,array}
\usepackage[hidelinks]{hyperref}

\newcommand{\F}{\mathbb F}
\newcommand{\Tr}{\operatorname{Tr}}
\newcommand{\tr}{\operatorname{tr}}
\newcommand{\N}{\mathcal N}
\newcommand{\ts}{\operatorname{t}}
\newtheorem{theorem}{Theorem}
\newtheorem{lemma}{Lemma}

\newtheorem{definition}{Definition}

\newtheorem{conjecture}{Conjecture}

\renewenvironment{abstract}
  {\par\noindent\textbf{Abstract:}\ \ignorespaces}
  {\unskip\par\medskip}

\renewenvironment{IEEEkeywords}
  {\par\noindent\textbf{Index Terms:}\ \ignorespaces}
  {\unskip\par\medskip}

\title{The APN Property of Two Families of functions over $\F_{2^{n}}$ from the Switching Construction}
	\author{{Chenmiao Shi, Longjiang Qu, Kangquan Li}\\
	\thanks{  Chenmiao Shi, Longjiang Qu, Kangquan Li are with the College of Science,
		National University of Defense Technology, Changsha, 410073, China.
		E-mail:  cmshi8300@163.com,   ljqu\_happy@hotmail.com, likangquan11@nudt.edu.cn.
		This work is supported by the National Cryptologic Science Fund of China under Grant (No. 2026NCSF02002) and the National Natural Science Foundation of China  (Nos. 12571579 and 12525115).
		{\emph{(Corresponding author: Kangquan Li)}}}
}

\begin{document}
\maketitle

\begin{abstract}
Let $m\geq5$ be odd with $3\nmid m$. Helleseth, Ichanska, and Kaleyski \cite{HIK}
conjectured that two norm-perturbed Budaghyan--Helleseth--Kaleyski functions
$F_2,F_3$ over $\F_{2^{2m}}$ are APN. In this paper, we prove this
conjecture. First, by using the basis $(1,\beta)$ of
$\F_{2^{2m}}$ over $\F_{2^m}$, we show that $F_2$ and $F_3$ are
EA-equivalent to two functions $\mathcal G_1$ and $\mathcal G_0$,
respectively. Their derivatives can be written as two trace equations for a
matrix $P\in\mathrm{GL}(2,\F_{2^m})$. We then prove that these two equations
have no solution. The proof uses only identities over finite fields, the
characteristic polynomial of a $2\times2$ matrix, and elementary properties
of Frobenius automorphisms. The repeated-root case and all parity cases are
considered separately. It follows that the derivative of each function has
kernel of size two in every nonzero direction, and hence both $F_2$ and
$F_3$ are APN for every admissible $m$.
\end{abstract}


\section{Introduction}

Let $n$ and $m$ be two positive integers. A mapping from $\F_{2^n}$ to
$\F_{2^m}$ is called an $(n,m)$-function, or a vectorial Boolean function.
Such mappings are used as substitution boxes in block ciphers and form the
main nonlinear component of many symmetric cryptosystems. Differential
cryptanalysis is one of the most efficient attacks on block ciphers
\cite{BihamShamir}, and differential uniformity was introduced to measure
the resistance of a function to this attack \cite{Nyberg}. When $n=m$, the
smallest possible differential uniformity is two, and the functions
attaining this bound are called almost perfect nonlinear (APN) functions.
Among the equivalence relations for vectorial Boolean functions, the
Carlet--Charpin--Zinoviev (CCZ) equivalence is the most general known one
preserving differential uniformity \cite{CCZ}. Consequently, constructing
APN functions that are CCZ-inequivalent to the known constructions is an
important but difficult problem. Prior to the recent switching families,
the known constructions were commonly summarized as six infinite families
of APN monomials, thirteen families in univariate polynomial form, nine
families in bivariate form, and two families in trivariate form; see
\cite{LiKaleyski} and the references therein.

Constructions of APN functions can be broadly divided into primary
constructions, which define APN functions directly, and secondary
constructions, which derive new APN functions from known ones. An important
secondary method is the switching construction of Edel and Pott
\cite{EdelPott}. Starting from an APN function $F$, it considers a
perturbation $F+cf$, where $c$ is a field element and $f$ is Boolean, and
seeks conditions under which the APN property is preserved. Although this
method has produced several sporadic APN functions, obtaining an infinite
family by switching is considerably more difficult: both the Boolean
perturbation and the proof of the APN property must work uniformly in the
field dimension. In 2009, by using the switching method to perturb the Gold function $x^3$ with a trace term, Budaghyan et al. constructed new infinite families of APN functions \cite{BCLswitch}.

Recently, in \cite{HIK}, Helleseth et al. applied the switching method in even dimension to the Budaghyan--Helleseth--Kaleyski (BHK) quadrinomials \cite{BHK} and obtained new infinite families by perturbing them with functions of the relative norm. This is the first construction of infinite families of APN functions obtained by applying the switching method to a non-power APN family. The
construction uses the Frobenius conjugates of the
relative norm $z^{2^m+1}$ as structured perturbation terms. Suitable
combinations of these conjugates give four families in even dimension,
organized into two switching pairs. In the first
pair, one construction is EA-equivalent to a previously known family that obtained by Li et al. \cite{Kangquan-Li-Yue-Zhou-2022},
whereas its switching neighbor was proved to be a new infinite family of
APN functions. For the second
pair, denoted by $F_2$ and $F_3$, extensive computations support the APN
property, and it was conjectured that both families are APN over
$\F_{2^{2m}}$ for every odd $m\geq5$ with $3\nmid m$.
\begin{conjecture}\label{conj}
\cite{HIK} Let $m\geq5$ be odd with $3\nmid m$, $n=2m$, $i=(m-2)^{-1}\pmod{n}$ and $k=\frac{m-3}{2}$. Let $\beta\in\F_4\setminus\F_2$ and
$$B(z)=z^3+\beta z^{2^i+1}+\beta^2 z^{3\cdot 2^m}+z^{(2^i+1)2^m}.$$
Then
\begin{align}
 &F_2(z)=B(z)+\beta\sum_{j=1}^{k}(z^{2^m+1})^{2^j}
 +\sum_{j=k+1}^{m-1}(z^{2^m+1})^{2^j},                                      \tag{1}\\
 F_3(z)&=B(z)+\beta^{2}z^{2^m+1}+\sum_{j=1}^{k}(z^{2^m+1})^{2^j}
 +\beta\sum_{j=k+1}^{m-1}(z^{2^m+1})^{2^j}.                                  \tag{2}
\end{align}
are APN over $\F_{2^n}$.
\end{conjecture}
The purpose of this
paper is to prove this conjecture. Our proof is based entirely on finite-field and elementary matrix
arguments. Using the basis $(1,\beta)$ of $\F_{2^{2m}}$ over $\F_{2^m}$,
we first show that $F_2$ and $F_3$ are EA-equivalent to two bivariate
functions $\mathcal G_1$ and $\mathcal G_0$, respectively. For a nonzero
derivative direction $A$ and a possible derivative zero $X$, we put
$T=\det(A\ X)$. When $T\ne0$, the derivative equations
are converted into two matrix equations. By studying the
traces, determinants, and characteristic polynomials of the
matrices involved, together with their behavior under
Frobenius automorphisms, we show that the two equations cannot
hold simultaneously. When $T=0$, a direct calculation shows that
$X\in\{0,A\}$. Hence every nonzero derivative has kernel of size two.

The remainder of this paper is organized as follows. Section~II recalls
the necessary definitions and finite-field facts. In Section~III, we derive
the EA-equivalent bivariate forms of $F_2$ and $F_3$. Section~IV presents the APN proof of $F_2$ and $F_3$, and Section~V concludes the paper.

\section{Preliminaries}

For positive integers $m\mid n$, the relative trace from $\F_{2^n}$ to
$\F_{2^m}$ is denoted by $\Tr_m^n$ and is defined by
\[
 \Tr_m^n(x)=\sum_{j=0}^{n/m-1}x^{2^{mj}}.
\]
In particular, $\Tr_1^n$ denotes the absolute trace on $\F_{2^n}$. For an
$(n,n)$-function $F$, let $\Delta_F(a,b)$ be the number of solutions
$x\in\F_{2^n}$ of
\[
 F(x+a)+F(x)=b.
\]
The differential uniformity of $F$ is
\[
 \delta_F=\max_{a\in\F_{2^n}^{*},\,b\in\F_{2^n}}\Delta_F(a,b).
\]
The function $F$ is APN if $\delta_F=2$.

For a quadratic function $F:\F_{2^n}\to \F_{2^n}$, define its symmetric derivative by
\[
 D_aF(x)=F(x+a)+F(x)+F(a)+F(0).
\]
For every $a\ne0$, $D_aF$ is $\F_2$-linear and
$\{0,a\}\subseteq\ker D_aF$. Therefore, $F$ is APN if and only if
\begin{equation}
 \ker D_aF=\{0,a\}                                                       \tag{3}
\end{equation}
for every $a\ne0$. 

We now recall several notions of equivalence between $(n,n)$-functions.
\begin{definition}\label{EA}
Two functions $F, G: \mathbb{F}_{2^{n}}\rightarrow \mathbb{F}_{2^{n}}$ are called extended affine equivalent (EA-equivalent), if there exist three affine functions $A_{1}, A_{2}, A_{3}: \mathbb{F}_{2^{n}}\rightarrow \mathbb{F}_{2^{n}}$, where $A_{1}, A_{2}$ are permutations, such that
$$F(A_{1}(x))=A_{2}(G(x))+A_{3}(x).$$
In particular, $F$, $G$ are called extended linear equivalent (EL-equivalent) if they are EA-equivalent and $A_{1}, A_{2}, A_{3}$ are linear mappings, $F$, $G$ are called  affine equivalent if they are EA-equivalent and $A_{3}=0$, $F$, $G$ are called linear equivalent if they are affine equivalent and $A_{1}, A_{2}$ are linear mappings.
\end{definition}

\begin{definition}\label{CCZ}
Two functions $F, G: \mathbb{F}_{2^{n}}\rightarrow \mathbb{F}_{2^{n}}$ are called Carlet-Charpin-Zinoviev equivalent (CCZ-equivalent), if their graphs are affine equivalent, that is, if there is an affine permutation $\mathcal{C}$ on $\mathbb{F}_{2^{n}}\times \mathbb{F}_{2^{n}}$ such that $\mathcal{C}({\rm \Gamma}_{F})={\rm \Gamma}_{G}$, where ${\rm \Gamma}_{F}=\{(x, F(x)): x\in \mathbb{F}_{2^{n}}\}$ is the graph of $F$.
\end{definition}

EA-equivalence implies CCZ-equivalence, and both equivalences preserve
differential uniformity and, in particular, the APN property.

Throughout the paper, let $m\geq5$ be odd with $3\nmid m$, and put
\[
q=2^m,\qquad K=\F_{q},\qquad L=\F_{q^2},\qquad \bar z=z^{q}.
\]
Let $\beta\in\F_4\setminus\F_2$ satisfy
$\beta^2+\beta+1=0$. Since $m$ is odd, we have
\[
 \F_4\cap K=\F_2,\qquad \bar\beta=\beta^q=\beta^2.
\]
Consequently, $(1,\beta)$ is a basis of $L$ over $K$. Denote
\[
N(z)=z\bar z, \qquad i=(m-2)^{-1}\pmod{2m}, \qquad k=\frac{m-3}{2}.
\]


We identify each power $2^r$ with the corresponding Frobenius
automorphism $z\mapsto z^{2^r}$ of $L$, and products and inverses of
such powers are understood as compositions and inverses of these
automorphisms. Put
\[
 s=\frac{m-1}{2},\qquad \sigma=2^s,\qquad
 \ts_s(z)=\sum_{j=0}^{s-1}z^{2^j}\quad(z\in K).
\]
As usual, $z^\sigma$ means $z^{2^s}$. The same notation is used entrywise
for vectors and matrices. The following identities will be used repeatedly.

For a field $E$ and an automorphism $\tau$ of $E$, the fixed field of
$\tau$ in $E$ is
\[
 \operatorname{Fix}_E(\tau)=\{z\in E:z^\tau=z\}.
\]

\begin{lemma}\label{lem:field}
Let $T\in K$, $b=\ts_s(T)$, and $\delta=\Tr_1^m(T)$. Then
\begin{equation}
 b^2+b=T^\sigma+T,\qquad
 b^\sigma+b=T^{\sigma^2}+\delta.                                       \tag{4}
\end{equation}
Moreover,
\begin{align*}
 \operatorname{Fix}_K(\sigma)&=\F_2,\\
 \operatorname{Fix}_{L}(\sigma)&=
 \begin{cases}
 \F_2,&s\text{ is odd},\\
 \F_4,&s\text{ is even}.
 \end{cases}
\end{align*}
Then the fixed field in $\F_{q^2}$ of
$q\sigma^{-1}=2^{s+1}$ is $\F_2$ when $s$ is even and $\F_4$ when $s$ is
odd. Finally,
\begin{equation}
 \sigma^m=\begin{cases}
 1,&s\text{ is even},\\
 q,&s\text{ is odd}
 \end{cases}
 \quad\text{on } L.                                                       \tag{5}
\end{equation}
\end{lemma}

\begin{IEEEproof}
The first identity in (4) follows by squaring the definition of $b$. Since
$m=2s+1$, adding $b$ and $b^\sigma$ gives all conjugates of $T$ except the
last one, and this proves the second identity. The fixed field of the map
$z\mapsto z^{2^r}$ on $\F_{2^d}$ is $\F_{2^{\gcd(r,d)}}$. Now
\begin{align*}
 \gcd(s,m)&=1,\\
 \gcd(s,2m)&=\gcd(s,2),\\
 \gcd(s+1,2m)&=\gcd(s+1,2),
\end{align*}
which gives all the fixed fields. Equation (5) follows from reducing $sm$
modulo $2m$.
\end{IEEEproof}

A one-dimensional $E$-subspace of $E^2$ is a set of the form
\[
 E(u,v)=\{\lambda(u,v):\lambda\in E\},
\]
where $(u,v)\ne(0,0)$. We shall use the following elementary observation. If such a line is
stable under a Frobenius automorphism $\tau$, then
it has a nonzero vector over $\operatorname{Fix}_E(\tau)$. Indeed, write the
line as $E(u,v)$. If $v\ne0$, its stability implies
$(u/v)^\tau=u/v$, so $(u/v,1)$ is such a vector. The case $v=0$ is clear.

A square matrix $R$ over a field is called \emph{nilpotent} if
$R^r=0$ for some positive integer $r$. If $R$ is a nonzero nilpotent
$2\times2$ matrix, then its minimal polynomial is a nonconstant divisor
of $X^r$ and has degree at most two. Since $R\ne0$, its minimal polynomial
cannot be $X$. It is therefore $X^2$, and hence $R^2=0$.

\begin{lemma}\label{lem:nilpotent}
Let $R_1,R_2$ be two nonzero nilpotent $2\times2$ matrices over a field of
characteristic two. If $\tr(R_1R_2)=0$, then
$\ker R_1=\ker R_2$.
\end{lemma}

\begin{IEEEproof}
Since $R_1^2=0$ and $R_1\ne0$, choose $e_2$ such that
$e_1=R_1e_2\ne0$. Then $R_1e_1=R_1^2e_2=0$, and $e_1,e_2$ are
linearly independent. Hence, with respect to the basis $(e_1,e_2)$,
the matrix of $R_1$ is
\[
 \begin{pmatrix}0&1\\0&0\end{pmatrix}.
\]
After applying this change of basis simultaneously to $R_1$ and $R_2$,
we may therefore assume that
\[
 R_1=\begin{pmatrix}0&1\\0&0\end{pmatrix}.
\]
Write
$R_2=\left(\begin{smallmatrix}u&v\\w&u\end{smallmatrix}\right)$; its two
diagonal entries are equal because $R_2$ is nilpotent and therefore has
trace zero. The
condition $\tr(R_1R_2)=0$ gives $w=0$, while
$\det R_2=u^2+vw=0$ gives $u=0$. Thus $R_2$ is a nonzero scalar multiple
of $R_1$, and the result follows.
\end{IEEEproof}

\section{An EA-equivalent bivariate form}

In this section, we show that the functions from $F_2$ and $F_3$ in Conjecture \ref{conj} are EA-equivalent to bivariate functions.

For a column $X=(x,y)^{\mathsf T}\in K^2$, define
\[
 M=\begin{pmatrix}0&1\\1&1\end{pmatrix},\qquad
 q_r(X)=\det(X,MX^{2^r}).                                                \tag{6}
\]
Then
\begin{equation}
 q_r(x,y)=x^{2^r+1}+xy^{2^r}+y^{2^r+1}.                                 \tag{7}
\end{equation}

\begin{lemma}\label{lem:EA}
For $e\in\F_2$, define $\mathcal G_e:K^2\to K^2$ by
\begin{equation}
 \mathcal G_e(x,y)=
 \left(
 \begin{array}{l}
 q_s(x,y)+\ts_s(xy)+e\Tr_1^m(xy),\\
 q_1(x,y)+xy+\Tr_1^m(xy)
 \end{array}
 \right).                                                               \tag{8}
\end{equation}
Then
\begin{equation}
 F_3\sim_{\mathrm{EA}}\mathcal G_0,\qquad
 F_2\sim_{\mathrm{EA}}\mathcal G_1.                                    \tag{9}
\end{equation}
\end{lemma}

\begin{IEEEproof}
Write the input of $L$ as $Z=x+\beta y$, where $x,y\in K$. Since
$m=2s+1$, we have $(m-2)s\equiv(-2)s=1-m\equiv1\pmod m$; hence
$i\equiv s\pmod m$. Moreover, $i$ is odd because it is the inverse of the
odd integer $m-2$ modulo $2m$. Therefore,
\[
 Z^{2^i}=x^\sigma+\beta^2y^\sigma.
\]
A direct calculation using $\beta^2+\beta+1=0$ gives
\begin{align}
 Z^3+\beta^2\bar Z^3&=\beta q_1(x,y),                                   \tag{10}\\
 \beta Z^{2^i+1}+\bar Z^{2^i+1}&=\beta^2q_s(x,y),                       \tag{11}\\
 N(Z)&=x^2+xy+y^2.                                                       \tag{12}
\end{align}
For every $j$, the difference
\[
 N(Z)^{2^j}-(xy)^{2^j}=x^{2^{j+1}}+y^{2^{j+1}}
\]
is an $\F_2$-linear function of $(x,y)$ and can be removed by an EA
transformation. Put $z=xy$. Modulo such linear terms, the perturbation in
$F_3$ is
\[
 \beta^2z+\sum_{j=1}^{s-1}z^{2^j}
 +\beta\sum_{j=s}^{m-1}z^{2^j}.
\]
Apply the invertible output map
\[
 u+\beta v\longmapsto (u,u+v).
\]
By (10)--(12), the BHK component $B(z)$ becomes $(q_s,q_1)$, and the displayed
perturbation becomes
\[
 \left(\sum_{j=0}^{s-1}z^{2^j},
 \sum_{j=1}^{m-1}z^{2^j}\right)
 =\bigl(\ts_s(z),z+\Tr_1^m(z)\bigr).
\]
This proves $F_3\sim_{\mathrm{EA}}\mathcal G_0$. The difference between the
two perturbations satisfies
\[
 F_2+F_3=\beta^2\Tr_1^m(N(Z)).
\]
Under the same output map, this adds $(\Tr_1^m(N(Z)),0)$. Moreover,
\[
 \Tr_1^m(N(Z))=\Tr_1^m(x)+\Tr_1^m(y)+\Tr_1^m(xy),
\]
the first two terms are linear and can again be removed. Thus
$F_2\sim_{\mathrm{EA}}\mathcal G_1$.
\end{IEEEproof}

Let $A,X\in K^2$ be two columns, put $P=(A\ X)$ and $T=\det P$. If
$T\ne0$, then $P\in\mathrm{GL}(2,K)$. From (6),
\begin{equation}
 D_Aq_r(X)=T\tr(P^{-1}MP^{2^r}).                                       \tag{13}
\end{equation}
Indeed, if $P^{-1}MP^{2^r}=
\left(\begin{smallmatrix}a&b\\c&d\end{smallmatrix}\right)$, then
\[
 MA^{2^r}=aA+cX,\qquad MX^{2^r}=bA+dX,
\]
and the polar part of $q_r$ is $(a+d)T$. The polar parts of $xy$,
$\Tr_1^m(xy)$, and $\ts_s(xy)$ are $T$, $\Tr_1^m(T)$, and $\ts_s(T)$,
respectively. Therefore, a nontrivial zero of a derivative of
$\mathcal G_e$ with $T\ne0$ gives a matrix $P$ satisfying the following system of two equations:
\begin{align}
 \tr(P^{-1}MP^\sigma)
 &=\frac{\ts_s(T)+e\Tr_1^m(T)}{T},                                     \tag{14}\\
 \tr(P^{-1}MP^2)
 &=1+\frac{\Tr_1^m(T)}{T}.                                              \tag{15}
\end{align}

\section{The APN property}
In this section, we prove the APN property of the functions from $F_2$ and $F_3$ in Conjecture \ref{conj}. We first prove that (14)--(15) have no solution.

\begin{lemma}\label{lem:matrix}
For every $e\in\F_2$, there is no matrix
$P\in\mathrm{GL}(2,K)$ satisfying (14)--(15).
\end{lemma}

\begin{IEEEproof}
Assume otherwise. Put
\begin{align*}
 b&=\ts_s(T),& \delta&=\Tr_1^m(T),\\
 S&=P^{-1}MP^\sigma,& U&=P^{-1}MP^2,
\end{align*}
and
\begin{equation}
 h=\frac{b+e\delta}{T},\qquad d=\frac{T^\sigma}{T}.                     \tag{16}
\end{equation}
Thus $\tr S=h$ and $\det S=d$.

Since $\sigma^2=2^{m-1}$ as an automorphism of $K$, $M^3=I$, and
$P^{2\sigma^2}=P$, we have
\[
 SS^\sigma=P^{-1}M^2P^{\sigma^2},\qquad
 U^{\sigma^2}=P^{-\sigma^2}MP^{2\sigma^2},
\]
and hence
\begin{equation}
 (SS^\sigma)U^{\sigma^2}=I.                                             \tag{17}
\end{equation}
For every invertible $2\times2$ matrix $V$,
$\tr(V^{-1})=\tr(V)/\det V$. By using (15), (17), and
$(T^{\sigma^2})^2=T$, we get
\begin{align}
 \tr(SS^\sigma)
 &=\det(SS^\sigma)(\tr U)^{\sigma^2}\notag\\
 &=\frac{T^{\sigma^2}+\delta}{T}.                                      \tag{18}
\end{align}
Here we used
$\det(SS^\sigma)=dd^\sigma=T^{\sigma^2}/T$ and
$(\tr U)^{\sigma^2}=1+\delta/T^{\sigma^2}$.
On the other hand, combining (18) with (4) and (16), we obtain
\begin{equation}
 \tr(SS^\sigma)=dh^\sigma+h
 =\frac{b^\sigma+b}{T}
 =\frac{T^{\sigma^2}+\delta}{T}.                                      \tag{19}
\end{equation}

The product
\begin{equation}
 \N_\sigma(S)=SS^\sigma\cdots S^{\sigma^{m-1}}
 =P^{-1}M^mP                                                            \tag{20}
\end{equation}
is obtained by using
$S^{\sigma^j}=P^{-\sigma^j}MP^{\sigma^{j+1}}$ and cancelling all
intermediate factors. Since $3\nmid m$,
$M^m$ is $M$ or $M^2$. Hence the matrix in (20) has characteristic
polynomial
\begin{equation}
 X^2+X+1.                                                               \tag{21}
\end{equation}
Because $m$ is odd, (21) is irreducible over $K$. In particular,
$\N_\sigma(S)$ has no one-dimensional invariant subspace over $K$.

We divide the remainder of the proof into two cases.

\textbf{Case 1: $h\ne0$.}

Put $c=b+e\delta$ and $u=T/c$. From $c^2+c=T^\sigma+T$, we obtain
\begin{equation}
 \frac{d}{h^2}=\frac{TT^\sigma}{c^2}=T+u+u^2,\qquad
 \Tr_1^m\left(\frac{d}{h^2}\right)=\delta.                             \tag{22}
\end{equation}
After substituting $X=hY$, the equation $X^2+hX+d=0$ becomes
$Y^2+Y=d/h^2$. The image of $Y\mapsto Y^2+Y$ on $K$ is precisely the
absolute-trace-zero subspace. Thus, by (22), the polynomial
$X^2+hX+d$ splits over $K$ if and only if $\delta=0$.
Let $E$ be its splitting field. Choose $v\in E$ satisfying
$v^2+v=T$. Since
\[
 \ts_s(v^2+v)=v^\sigma+v,
\]
we have $v^\sigma+v=b$. Set
\[
 w=u+v,\qquad \ell=hw,\qquad \ell'=\ell+h.
\]
By (22), $\ell$ and $\ell'$ are the two roots of $X^2+hX+d$.

Let
\[
 A_0=\ell\ell^\sigma+\ell'{\ell'}^\sigma,\qquad
 B_0=\ell{\ell'}^\sigma+\ell'\ell^\sigma.
\]
Then $A_0+B_0=hh^\sigma$. Moreover,
\begin{equation}
 dh^\sigma+h=A_0+e\delta hh^\sigma.                                    \tag{23}
\end{equation}
Indeed, 
\begin{align*}
\frac{dh^\sigma+h+A_0}{hh^\sigma}&=w+w^\sigma+1+du+u^\sigma\\
 &=u+b+1+du\\
 &=\frac{T+T^\sigma}{c}+b+1\\
 &=e\delta.
\end{align*}

Let $Q\in\mathrm{GL}(2,E)$ be an eigenvector matrix of $S$, and write
\[
 S=Q\operatorname{diag}(\ell,\ell')Q^{-1},\qquad
 C=Q^{-1}Q^\sigma=
 \begin{pmatrix}\alpha&\gamma_0\\ \gamma_1&\varepsilon\end{pmatrix}.
\]
Put $\Delta=\alpha\varepsilon+\gamma_0\gamma_1$. A direct multiplication
gives
\begin{equation}
 \tr(SS^\sigma)
 =A_0+\frac{\gamma_0\gamma_1}{\Delta}(A_0+B_0).                         \tag{24}
\end{equation}
Comparing (19), (23), and (24), we obtain
\begin{equation}
 \frac{\gamma_0\gamma_1}{\Delta}=e\delta.                              \tag{25}
\end{equation}

\textbf{Subcase 1.1: $\delta=0$.}

Both eigenvalues and eigenlines of $S$ are defined over $K$. Equation (25)
shows that $C$ is upper or lower triangular. Hence at least one eigenline is
stable under $\sigma$. The semilinear map
\begin{equation}
 \Phi(z)=Sz^\sigma                                                       \tag{26}
\end{equation}
preserves this $K$-line. Its $m$-th iterate, whose matrix is
$\N_\sigma(S)$, preserves the same line, contradicting (20)--(21).

\textbf{Subcase 1.2: $\delta=1$.} 

Then $E=\F_{q^2}$, where $q=2^m$, and the
$q$-Frobenius interchanges the two eigenlines. Choose the two columns of $Q$
as a conjugate pair. Then
\[
 Q^q=QJ,\qquad J=\begin{pmatrix}0&1\\1&0\end{pmatrix}.
\]
Since the $q$- and $\sigma$-Frobenius maps commute, we have
$JC^q=CJ$. Thus
\begin{equation}
 \alpha^q=\varepsilon,\qquad \gamma_0^q=\gamma_1.                       \tag{27}
\end{equation}

\textbf{Subcase 1.2.1: $e=0$.}

By (25) and (27), $\gamma_0=\gamma_1=0$. Thus both eigenlines are stable
under $\sigma$. If $s$ is odd, Lemma~\ref{lem:field} and the elementary
fixed-line observation imply that both lines are defined over $\F_2$. This
contradicts the irreducibility of $X^2+hX+d$ over $K$: a $K$-rational
eigenline of the matrix $S\in\mathrm{GL}(2,K)$ has an eigenvalue in $K$.

Let $s$ be even. Since
$\Tr_1^m(T+1)=\delta+\Tr_1^m(1)=0$, choose $a\in K$ satisfying
$a^2+a=T+1$. Let $\omega\in\F_4\setminus\F_2$ and put
$\rho=a+\omega$. Then
\[
 T=\rho\rho^q,\qquad \omega^\sigma=\omega,\qquad
 b=\ts_s(T)=a^\sigma+a.
\]
Consequently,
\begin{equation}
 \lambda=\frac{\rho^\sigma}{\rho}                                     \tag{28}
\end{equation}
satisfies
\[
 \lambda+\lambda^q
 =\frac{\rho^\sigma\rho^q+\rho^{\sigma q}\rho}{T}
 =\frac{a^\sigma+a}{T}=h,
 \qquad
 \lambda\lambda^q=\frac{T^\sigma}{T}=d.
\]
Hence it is one of $\ell,\ell'$. The product of its $\sigma$-conjugates is
\[
 \lambda\lambda^\sigma\cdots\lambda^{\sigma^{m-1}}
 =\frac{\rho^{\sigma^m}}{\rho}=1
\]
by (5); the same holds for $\lambda^q$. The two eigenlines have vectors over
$\F_4=\operatorname{Fix}_E(\sigma)$. Using these vectors as the columns of
$Q$ gives $Q^\sigma=Q$, and therefore
\[
 Q^{-1}\N_\sigma(S)Q=
 \operatorname{diag}\left(
 \prod_{j=0}^{m-1}\lambda^{\sigma^j},
 \prod_{j=0}^{m-1}(\lambda^q)^{\sigma^j}\right)=I.
\]
This contradicts (20)--(21).

\textbf{Subcase 1.2.2: $e=1$.}

Equation (25) gives
$\gamma_0\gamma_1=\Delta=\alpha\varepsilon+\gamma_0\gamma_1$, so
$\alpha\varepsilon=0$. By (27), $\alpha=\varepsilon=0$. Therefore,
$\sigma$ interchanges the two eigenlines. If $s$ is even, the automorphism
$q\sigma^{-1}=2^{s+1}$ fixes both lines. Its fixed field is $\F_2$ by
Lemma~\ref{lem:field}, so the two lines are defined over $K$, a
contradiction.

It remains to consider odd $s$. Choose $a\in K$ satisfying
$a^2+a=T+1$, and put $\rho=a+\omega$, where
$\omega\in\F_4\setminus\F_2$. Now $\omega^\sigma=\omega^2$ and
\[
 b=\ts_s(T)=a^\sigma+a+1,\qquad c=b+1=a^\sigma+a.
\]
The element
\begin{equation}
 \lambda=\frac{\rho^\sigma}{\rho^q}                                   \tag{29}
\end{equation}
satisfies
\[
 \lambda+\lambda^q
 =\frac{\rho^\sigma\rho+\rho^{\sigma q}\rho^q}{T}
 =\frac{a^\sigma+a}{T}=h,
 \qquad
 \lambda\lambda^q=\frac{T^\sigma}{T}=d.
\]
Thus $\lambda,\lambda^q$ are the eigenvalues of $S$. The two eigenlines are
fixed by $q\sigma^{-1}$, whose fixed field is $\F_4$. Choose their
eigenvector matrix $Q\in\mathrm{GL}(2,\F_4)$ as follows: choose a nonzero
vector on the first line and take its $q$-th power as the second column.
Then $Q^q=QJ$. Since both
$m$ and $s$ are odd, the $q$- and $\sigma$-Frobenius maps act as squaring on
$\F_4$, and hence $Q^\sigma=QJ$ as well. In the $Q$-coordinates, the map
(26) has matrix
\[
 A=\operatorname{diag}(\lambda,\lambda^q)J.
\]
The product $R=AA^\sigma\cdots A^{\sigma^{m-1}}$ is anti-diagonal. Its
upper-right entry is
\begin{equation}
 \Lambda=
 \prod_{\substack{0\le j<m\\j\text{ even}}}
 \frac{\rho^{\sigma^{j+1}}}{\rho^{q\sigma^j}}
 \prod_{\substack{0\le j<m\\j\text{ odd}}}
 \frac{\rho^{q\sigma^{j+1}}}{\rho^{\sigma^j}}.                         \tag{30}
\end{equation}
By (5), $q=\sigma^m$ on $E$. The numerator and denominator in (30) contain
exactly the same conjugates
\[
 \rho^\sigma,\rho^{\sigma^3},\ldots,\rho^{\sigma^{2m-1}},
\]
because the numerator exponents are
\[
 \{j+1:j\ {\rm even}\}\cup\{m+j+1:j\ {\rm odd}\},
\]
whereas the denominator exponents are
\[
 \{m+j:j\ {\rm even}\}\cup\{j:j\ {\rm odd}\};
\]
where $0\leq j<m$ in all four sets. Both sets are exactly the odd residues
modulo $2m$. Hence $\Lambda=1$.
The lower-left entry is $\Lambda^q=1$, and consequently $R=J$.
For completeness, if $z=Qy$, then the two descriptions of the $m$-th
iterate of (26) give
\[
 \Phi^m(Qy)=QRy^q=\N_\sigma(S)Q^qy^q.
\]
Since $Q^q=QJ$, it follows that
\[
 Q^{-1}\N_\sigma(S)Q=RJ=I,
\]
contradicting (20).

\textbf{Case 2: $h=0$.}

Then $b+e\delta=0$. By (4), $T^\sigma=T$. Since
$\operatorname{Fix}_K(\sigma)=\F_2$ and $T\ne0$, we have $T=1$ and $d=1$.
The characteristic polynomial of $S$ is $(X+1)^2$, so write
$S=I+R$, where $R^2=0$. If $R=0$, the left-hand side of (20) is $I$, which
is impossible. Hence $R$ is a nonzero rank-one nilpotent matrix. From (19),
\[
 0=\tr(SS^\sigma)=\tr(RR^\sigma).
\]
Indeed, after expanding $(I+R)(I+R^\sigma)$, the traces of $I$, $R$, and
$R^\sigma$ all vanish in characteristic two.
By Lemma~\ref{lem:nilpotent}, $R$ and $R^\sigma$ have the same kernel.
Since $\ker R^\sigma=(\ker R)^\sigma$, the $K$-line $\ker R$ is stable
under $\sigma$. It is preserved by (26), and consequently by its $m$-th
iterate $\N_\sigma(S)$. This again contradicts (20)--(21).

All cases lead to a contradiction, and the proof is complete.
\end{IEEEproof}

We now combining Lemma \ref{lem:matrix} to prove the APN property of the functions from $F_2$ and $F_3$.
\begin{theorem}\label{thm:main}
Let $m\geq5$ be odd with $3\nmid m$. The two functions $F_2$ and $F_3$
defined in (1)--(2) are APN over $\F_{2^{2m}}$.
\end{theorem}

\begin{IEEEproof}
By Lemma~\ref{lem:EA}, it is enough to prove that $\mathcal G_e$ is APN for
$e=0,1$. Let $A=(a,b)^{\mathsf T}\in K^2\setminus\{0\}$ be a derivative
direction and let $X=(x,y)^{\mathsf T}$ be in the kernel of its symmetric
derivative. Put $T=\det(A\ X)$.

If $T\ne0$, equations (13)--(15) give a matrix
$P=(A\ X)\in\mathrm{GL}(2,K)$ satisfying (14)--(15), which is impossible by
Lemma~\ref{lem:matrix}.

Now let $T=0$. Then $X=\mu A$ for some $\mu\in K$. The second coordinate of
the derivative of $\mathcal G_e$ is
\begin{equation}
 (\mu^2+\mu)q_1(A).                                                      \tag{31}
\end{equation}
We claim that $q_1(A)\ne0$. If $b=0$, this is clear. If $b\ne0$ and
$q_1(A)=0$, then $a/b$ is a root of $Z^3+Z+1$. This polynomial is
irreducible over $\F_2$ and its roots belong to $\F_8\setminus\F_2$.
However,
\[
 K\cap\F_8=\F_{2^{\gcd(m,3)}}=\F_2,
\]
which is a contradiction. Thus (31) vanishes only when
$\mu\in\F_2$, and therefore $X=0$ or $X=A$.

We have proved that the kernel of every nonzero derivative is exactly
$\{0,A\}$. Hence $\mathcal G_0$ and $\mathcal G_1$ are APN. By (9), the
same is true for $F_3$ and $F_2$, respectively.
\end{IEEEproof}

\section{Conclusion}

We have proved that the two conjectured families $F_2$ and $F_3$ are APN
over $\F_{2^{2m}}$ for every odd $m\geq5$ with $3\nmid m$. The main step is
the EA-equivalent form (8), which transforms every possible nontrivial
derivative zero with $T\neq 0$ into the two matrix equations (14)--(15). The local identity
(19) controls the Frobenius action on the two eigenlines of $S$, while the
global product (20) has no invariant line over $K$. The resulting
contradiction excludes all nontrivial derivative zeros and completes the
proof of Conjecture 1 in \cite{HIK}.


\begin{thebibliography}{99}
\normalsize


\bibitem{BihamShamir}
E.~Biham and A.~Shamir,
``Differential cryptanalysis of DES-like cryptosystems,''
\emph{J. Cryptology}, vol.~4, no.~1, pp.~3--72, 1991.

\bibitem{BCLswitch}
L.~Budaghyan, C.~Carlet, and G.~Leander,
``Constructing new APN functions from known ones,''
\emph{Finite Fields Appl.}, vol.~15, no.~2, pp.~150--159, 2009.

\bibitem{CCZ}
C.~Carlet, P.~Charpin, and V.~Zinoviev,
``Codes, bent functions and permutations suitable for DES-like
cryptosystems,'' \emph{Des. Codes Cryptogr.}, vol.~15, no.~2,
pp.~125--156, 1998.

\bibitem{EdelPott}
Y.~Edel and A.~Pott,
``A new almost perfect nonlinear function which is not quadratic,''
IACR Cryptology ePrint Archive, Paper 2008/313, 2008.

\bibitem{HIK}
T.~Helleseth, N.~Ichanska, and N.~S.~Kaleyski,
``New infinite families of APN functions from the switching construction in
even dimension,'' IACR Cryptology ePrint Archive, Paper 2026/2401, 2026.

\bibitem{BHK}
L.~Budaghyan, T.~Helleseth, and N.~S.~Kaleyski,
``A new family of APN quadrinomials,'' \emph{IEEE Trans. Inf. Theory},
vol.~66, no.~11, pp.~7081--7087, Nov.~2020.

\bibitem{LiKaleyski}
K.~Li and N.~S.~Kaleyski,
``Two new infinite families of APN functions in trivariate form,''
\emph{IEEE Trans. Inf. Theory}, vol.~70, no.~2, pp.~1436--1452,
Feb.~2024.


\bibitem{Kangquan-Li-Yue-Zhou-2022}
K.~Li, Y.~Zhou, C.~Li, and L.~Qu,
``Two new infinite families of quadratic APN functions,''
\emph{IEEE Trans. Inf. Theory}, vol.~68, no.~7, pp.~4761--4769,
Feb.~2022.

\bibitem{Nyberg}
K.~Nyberg,
``Differentially uniform mappings for cryptography,'' in
\emph{Advances in Cryptology---EUROCRYPT'93}, Lecture Notes in Computer
Science, vol.~765. Berlin, Germany: Springer, 1994, pp.~55--64.
\end{thebibliography}
\end{document}